# Experimental results of the laserwire emittance scanner for LINAC4 at CERN

T. Hofmann[a,b,c]*, G.E. Boorman[b], A. Bosco[b], E. Bravin[a], S.M. Gibson[b], K.O. Kruchinin[b], U. Raich[a], F. Roncarolo[a], F. Zocca[a]

[a]*CERN, Geneva, 1211, Switzerland*
[b]*John Adams Institute at Royal Holloway, University of London, Egham, TW20 0EX, United Kingdom*
[c]*Friedrich-Alexander-University, Erlangen, 91054, Germany*

## Abstract

Within the framework of the LHC Injector Upgrade (LIU), the new LINAC4 is currently being commissioned to replace the existing LINAC2 proton source at CERN. After the expected completion at the end of 2016, the LINAC4 will accelerate H$^-$ ions to 160 MeV. To measure the transverse emittance of the H$^-$ beam, a method based on photo-detachment is proposed. This system will operate using a pulsed laser with light delivered via an optical fibre and subsequently focused through a thin slice of the H$^-$ beam. The laser photons have sufficient energy to detach the outer electron and create H$^0$/e$^-$ pairs. In a downstream dipole, the created H$^0$ particles are separated from the unstripped H$^-$ ions and their distribution is measured with a dedicated detector. By scanning the focused laser across the H$^-$ beam, the transverse emittance of the H$^-$ beam can be reconstructed. This paper will first discuss the concept, design and simulations of the laser emittance scanner and then present results from a prototype system used during the 12 MeV commissioning of the LINAC4.

## Introduction

The photo-detachment process for hydrogen ions has been known since 1959 [1]. By launching H$^-$ ion based linear accelerators (linacs) this technique was firstly applied in order to measure parameters of the H$^-$ beam [2]. For today's modern high current linacs this technique offers numerous benefits. Due to its non-invasive nature, beam parameters can be measured parasitically and since no mechanical parts are needed to intercept the particle beam, the risk of overheated and damaged components is eliminated. Since the reduction of downtimes is a major aim for any accelerator, this non-destructive instrument can contribute to maximize the efficiency.

* Thomas Hofmann. Tel.: +41-22-76-77130.
E-mail address: thomas.hofmann@cern.ch

The advantages are particularly relevant for the LINAC4 which is due to deliver high brightness beams reliably to the LHC injector complex. The declared aim of the HL-LHC, to increase the integrated luminosity by a factor of ten [3], is directly dependent on the LINAC4 performance and reliability. The increased final energy of 160 MeV creates further challenges for the beam instrumentation. Due to the long range of protons at this energy (e.g. > 20 cm in graphite) conventional methods such as slits or scrapers cannot be used. By using a laser as a 'slit' not only is the range problem solved, but the laser can be adjusted such that just a tiny fraction of $H^-$ ions become neutralized $H^0$. This makes the laserwire principle truly non-invasive. A comprehensive overview of the laserwire system and the integration at LINAC4 can be found at [4], [5].

In this paper the emphasis is put on measurement results obtained while commissioning such a laserwire system on a 12 MeV $H^-$ beam. Since the principle of this technique has been already demonstrated in other facilities [6] [7] [8], the focus was on making advances in key parameters of the system. In this regard a low-power laser was used which made it possible to design a simple fibre-based system, which runs very reliably and has imperceptible impact on the ion beam.

**Instrument design**

*1.1. LINAC4*

LINAC4 is the first step in the LHC Injector Upgrade (LIU) program which is essential for delivering the high brightness beams for the HL-LHC upgrade. After determining the principal machine parameters in a technical design report in 2006 [9], the civil engineering was completed and the machine is currently in the commissioning phase. Table 1 and Figure 1 provide an overview on the machine's essential parameters.

Table 1. Principal LINAC4 parameters

| Parameter | Value |
| --- | --- |
| Overall linac length | 90 m |
| Output energy | 160 MeV |
| Bunch frequency | 352 MHz |
| Beam pulse length | 400 µs |
| Beam pulse repetition rate | 0.83 Hz |
| Average pulse current | 40 mA |
| Nominal transverse emittance | 0.4 π mm mrad |

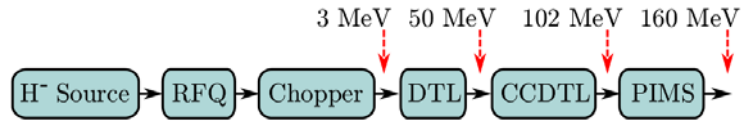

Figure 1. LINAC4 block diagram. After generating the ion beam in a caesiated surface source the RFQ first accelerates and then generates a bunched beam at an energy of 3 MeV. The chopper selects a temporal slice of the beam for further acceleration. DTL, CCDTL and PIMS are the drift tubes linacs (DTL) required to accelerate the beam to the final energy of 160 MeV.

The determination of the transverse emittance at the end of the linac is essential in order to avoid high losses when the beam is injected into the PS-Booster [10]. Therefore the 160 MeV region is the envisaged final location for the proposed laserwire instrument. To advance with the instrument's design while the commissioning of LINAC4 is still underway, the phases at 3 MeV and 12 MeV (after the first DTL tank) were used to test the laserwire prototype.

*1.2. Instrument concept*

Due to the extremely small binding energy of the outer electron of 0.75 eV, the $H^-$ ion can be neutralized rather easily. Collisions with residual gas atoms [4] or strong magnetic fields in the Tesla range can detach one electron. The principle of our instrument is based on electron detachment due to collisions with photons. This process has a large cross-section of $>3 \times 10^{-17}$ cm$^2$ in the range between 500 nm and 1200 nm photon wavelength [1].

To sample the transverse phase space of an ion beam, a common method is to select thin slices of the beam and measure its angular distribution, known as the slit/grid method [11]. In analogy to this principle, Figure 2 shows a schematic of the laserwire, which in contrast to the slit/grid method, is completely non-destructive.

## 1.3. Laser system

Since the laser system was described in detail in previous publications [4], [5] it will be summarised here. The general parameters of the chosen working point of the laser source are listed in Table 2. The low pulse energy in comparison to other laserwire systems makes it possible to deliver the laser light to the particle beam via a long optical fibre (10 m was already achieved during the experiments completed at 3 and 12 MeV and a longer fibre will be used for the final system). This greatly reduces the complexity of the overall laser delivery system.

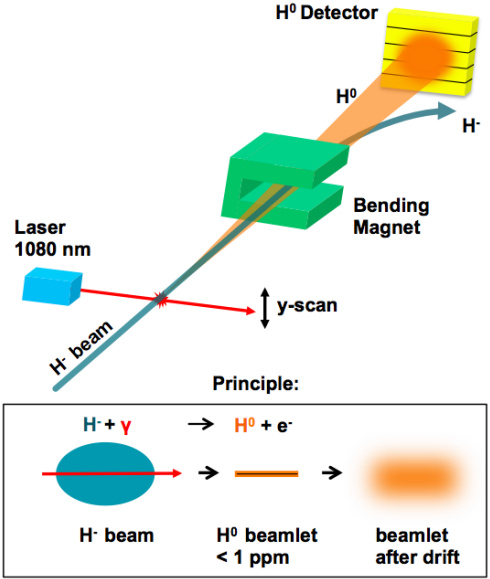

Figure 2. Concept for laserwire emittance measurement. A focused laser crosses the H$^-$ beam and detaches electrons from the ions. The resulting neutral H$^0$ atoms drift unperturbed towards a detector while the H$^-$ ions are deflected in a downstream bending magnet. By measuring the H$^0$ profiles in the detection plane, the angular distribution of the beam is gained. A scan of the laser through the H$^-$ beam allows sampling the transverse phase space. [12]

Table 2. Laser parameters

| Parameter | Value |
| --- | --- |
| Wavelength | 1080 nm |
| Pulse energy | 154 µJ |
| Pulse length (FWHM) | 80 ns |
| Pulse freq. | 60 kHz |
| $M^2$ | 1.8 |

The laser is subsequently focused into the vacuum vessel with a diameter of approximately 150 µm. Due to the quasi-monomode beam quality ($M^2 = 1.8$), the laser beam diameter remains almost constant when colliding with the millimetre-size particle beam. Vertical scanning of the laser is performed by a remote controlled stage. A CCD camera and a fast photodiode are used to monitor the laser beam quality.

## 1.4. H$^0$ Detection system

For the measurements at 12 MeV, the 20 mm x 20 mm polycrystalline diamond detector with 5 strip channels already exploited during the 3 MeV experiments [13] was used to detect the neutralized H$^0$ atoms. Contrary to the 3 MeV case, the 12 MeV H$^0$ atoms fully traverse the 500 µm thick diamond substrate. This eliminates problems with piled up charge inside the diamond and greatly increases the signal level. A simulation of the ionization inside the diamond was performed using the software package SRIM [14]. It was found that one H$^0$ loses 9.3 MeV when traversing the diamond. The created charge readout at the diamond can be calculated as:

$$Q_{Diamond} = n_{H^0} \cdot e \cdot \frac{E_{ionize}}{E_{Gen}} \cdot CCE \qquad (1)$$

where $n_{H0}$ is the number of H$^0$ hitting the diamond, $E_{ionize}$ the energy that one H$^0$ loses in the diamond and $E_{Gen}$ the Generation Energy, which is 13.1 eV for diamond [15]. Depending on the diamond material quality the Charge Collection Efficiency (CCE) can vary significantly. For the detector used in the measurements a CCE between 10% and 20% was assumed [16]. Using this value, a simulation was executed to estimate the created charge taking into account the power of the laser as presented in Table 2, the beam dynamics and the size of a detector strip channel (3.5 mm x 18 mm). A charge of up to $2 \times 10^{-9}$ C was found to be created by one laser pulse in one detector channel.

Such a charge creates a voltage in the order of 1 V (across a 50 Ohm termination) for the duration of a laser pulse, and the signal can therefore be digitized without pre-amplification.

## System characterization

### 1.1. *Signal amplitude at diamond detector*

The signals received from the diamond detector were compared to simulations to verify that the laser and detector systems worked as expected. The left plot of Figure 3 shows the signal from the diamond detector caused by one laser pulse. The background level is a result of $H^0$ generated by interactions with residual gas atoms [4]. To determine the charge created by the laser interaction, this background was subtracted and the resulting distribution was then integrated. On the right side of Figure 3, the resulting charge values are plotted in the vertical phase space.

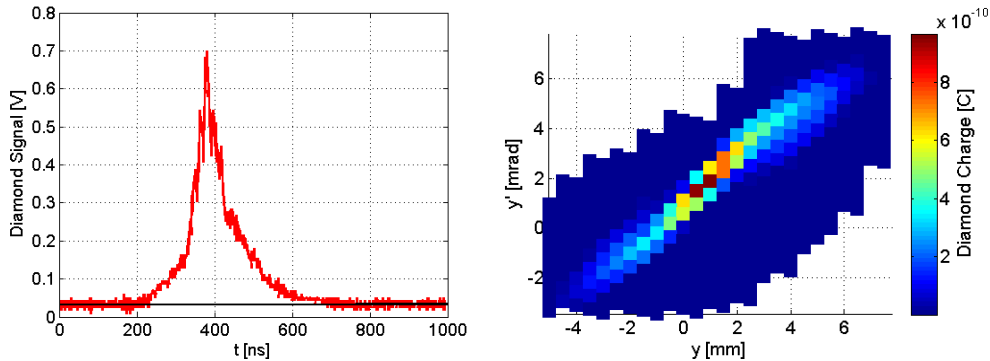

Figure 3. Left: Signal on diamond by one laser pulse; Right: Integrated charge of diamond detector in phase space

The measured charge values are approximately a factor of 2 lower than expected. Taking into account the uncertainties in the simulation due to the indefinite CCE value of the detector, the agreement can be considered as acceptable. It must be also noted that the CCE changes due to radiation damage and therefore the amount of charge collected also varied during the two month measurement campaign.

### 1.2. *Signal linearity along beam pulse*

In order to determine the linearity of the diamond detector response along the ion beam pulse, a comparison with a beam current transformer (BCT) was performed. Special attention was put on this test, since the previous results using the 3 MeV beam [13] showed rather non-linear behaviour.

The signal from the BCT situated immediately after the laser interaction point (IP) was measured for the duration of the phase space scan. The signal from the diamond detector was recorded as shown in Figure 4. A whole phase space scan was made for each detector channel. The phase space was then integrated for comparison with the averaged BCT signal during the scan period, as shown in Figure 5.

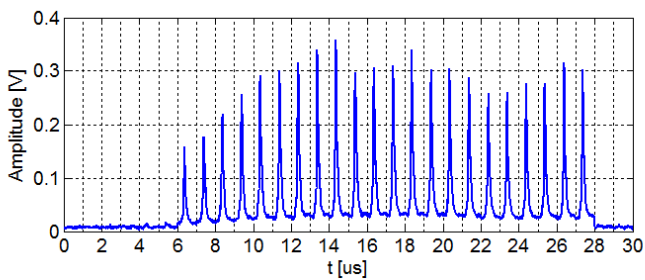 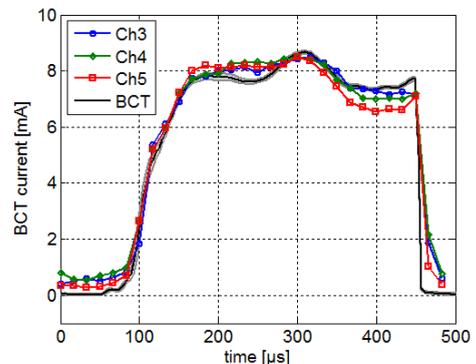

Figure 4. Diamond signal segments along 400 µs H⁻ ion pulse. The laser runs at 60 kHz and creates 24 signal peaks in the signal of the diamond detector. For each peak, a 1 µs time interval is recorded. The plot shows these signal intervals stitched together for the duration of the ion pulse.

Figure 5. Comparison of the averaged beam current with the signals of 3 channels of the diamond detector. The diamond signals were normalized to compare with the BCT signal.

The plot shows that the behaviour of each of the three diamond detector channels is very similar. The agreement between each channel and the averaged BCT signal is within 10% for each sample of the pulse. The residual disagreement is the result of a small number of particles bypassing the diamond detector. This is a significant improvement over the 3 MeV measurements, for which the diamond signal dropped dramatically in the second half of the beam pulse, which was caused by the pile-up of implanted H⁻ ions in the diamond bulk material.

### 1.3. *Homogeneity of diamond detector*

The signals from each of the diamond channels were compared to check the homogeneity of the polycrystalline diamond material. Since the DAQ system was still in development, the system was only capable of recording the signals from three out of the five detector channels. The focused laser was kept at a fixed position, and the detector was moved to record each detector channel at the same position in phase space.

The curves in Figure 6 correspond to the integrated charge per laser pulse in each segment. The difference between channels is well within their error bars, which correspond to the standard deviation (SD) of successive recordings. The increase in error during the rising and falling edge is due to the missing synchronization of the 60 kHz laser with the ion beam, which introduced a 17 µs jitter on the signal from the diamond detector.

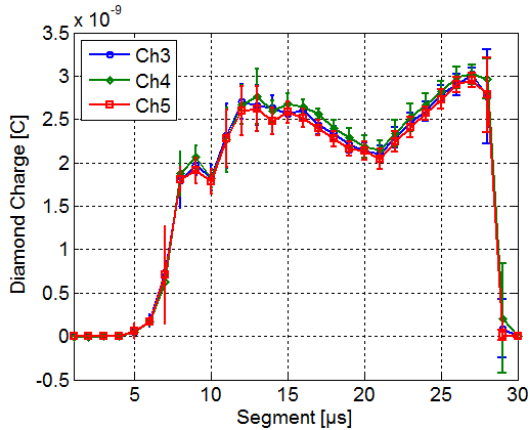

Figure 6. Signal charge on three detector channels. Recorded with laser at fixed position and each detector channel moved to common location.

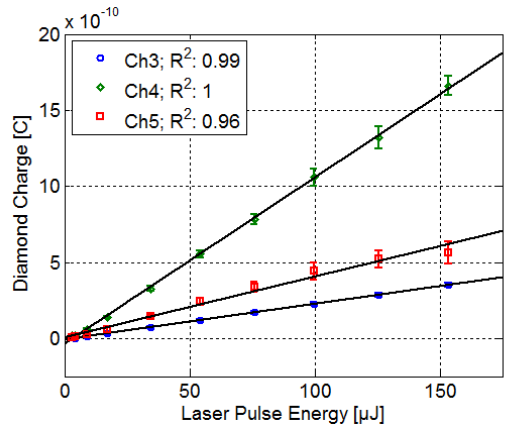

Figure 7. Integrated charge of diamond detector channels while scanning the laser pulse energy

### 1.4. *Signal linearity versus laser pulse energy*

A key aspect in determining the beam profile and the emittance is the linearity of the diamond detector with respect to the impinging particles. It is difficult to change the LINAC4 ion beam current without changing the beam divergence and size, therefore a different approach was chosen. Assuming a linear relation between laser pulse energy and stripped H⁻ ions, the linearity of the diamond detector can be determined by varying the laser power.

The laser pulse energy at the end of the fibre was measured and the laser focus was positioned at the centre of the H⁻ beam. The results with the detector at a fixed position are shown in Figure 7. The incline of the slopes are different due to the different positions of the detector channels. However, the linear fit of the data always shows a coefficient of determination ($R^2$) greater than 0.96, showing the signal distortion due to non-linearities is negligible.

**Data analysis and results**

1.5. *Temporal beam properties*

The measurement campaign was mainly focused on sampling the phase space and then reconstructing the transverse beam emittance and profile. However, by using a high bandwidth detection system temporal beam properties could also be measured. Figure 8 shows the laser pulse recorded before delivery to the beam pipe and the signal from the diamond detector. The similar shape shows clearly that the signal from the diamond detector originates from $H^0$ atoms created by the laser pulse. The delay of the 2 signals of 194 ns (SD, 4 ns) consists of three parts: the time of the laser delivery to the IP with the particle beam (51 ns); the time-of-flight (TOF) of the particle from IP to the diamond detector (71 ns); and 72 ns delay in the cable from the diamond detector to the DAQ system. This system cannot replace a dedicated system for TOF-measurement but it gives a hint regarding the energy of the $H^-$ ions.

A further observation was made by performing a Fourier-Transform of the diamond detector signal. In Figure 9 the spectra of the signals in the previous figure are plotted. The strongest frequency components in the lower frequency range are very similar when comparing the laser pulse recorded at the PD and the diamond. In the high-frequency range a clear signal from the diamond is recorded at exactly 352 MHz, the bunch frequency of the LINAC4. This illustrates the potential of the diamond detector regarding measurements with highest time resolution.

The absence of the 352 MHz peak in the background signal is also well understood. The $H^0$ particles in the background are generated by collisions with residual gas atoms all along the accelerator, each with varied kinetic energy and the arriving particles are therefore expected to be completely de-bunched.

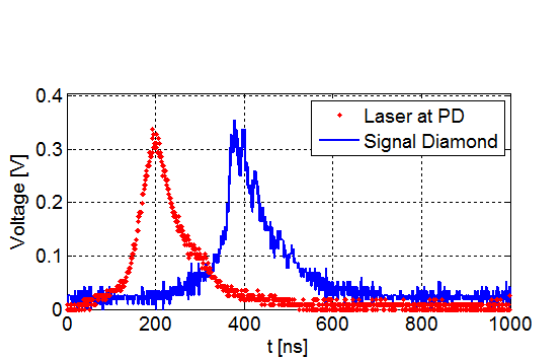

Figure 8. Laser pulse recorded with fast photodiode (PD) and signal of diamond detector

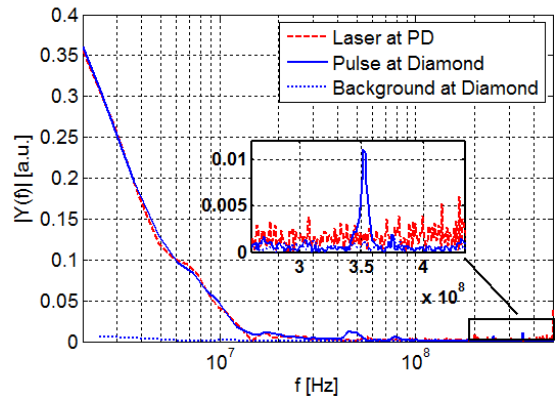

Figure 9. Fourier transformation of the laser pulse at the PD, the pulse at the diamond detector and the constant background arriving at the diamond detector. Zoom: Peak at 352 MHz bunch freq.

1.6. *Transverse phase space and beam profile*

As for the 3 MeV case, a system for sampling only the vertical phase space was developed for the 12 MeV experiments.

Just before the set of emittance measurements started, a fault in the LINAC4 machine interlock system allowed a full $H^-$ pulse to hit the diamond detector. The resulting radiation damage left only one diamond channel surviving with sufficient CCE. Consequently the emittance scan consisted of vertically scanning the detector through the whole $H^0$ beamlet in steps of 1.8 mm. The detector scan was repeated for each laser position, changed in steps of 0.5 mm.

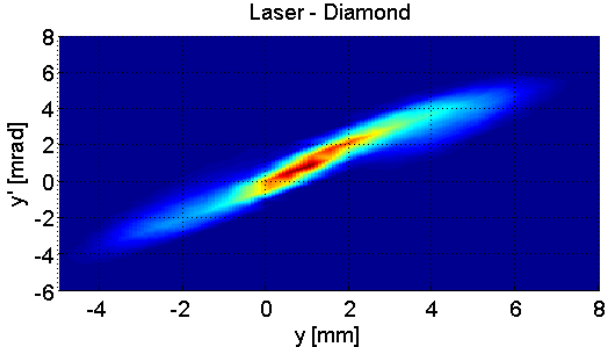

Figure 10. Phase space sampled with the laser/diamond system

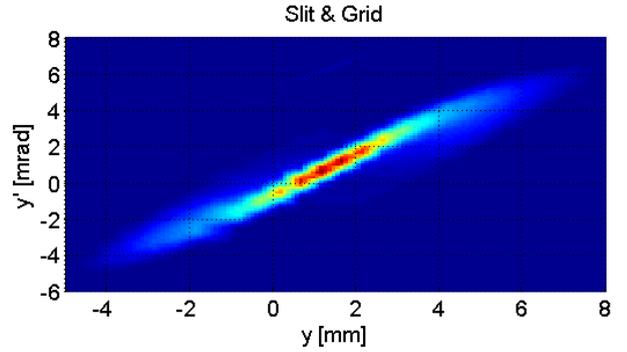

Figure 11. Phase space sampled with the slit/grid method.

The raw data (of the kind shown in Figure 4) was analysed in three steps. The detector signal was first low-pass filtered to remove high frequency noise. Then a linear fit to the background level was made (see left of Figure 3) and subtracted from the signal, before summing the charge of the resulting signal. The obtained values in vertical phase space are shown in Figure 10.

To evaluate the precision of the novel laserwire instrument, the beam was sampled also with a conventional slit/grid instrument. Figure 11 shows the result of the reference measurement using the same H⁻ beam parameters.

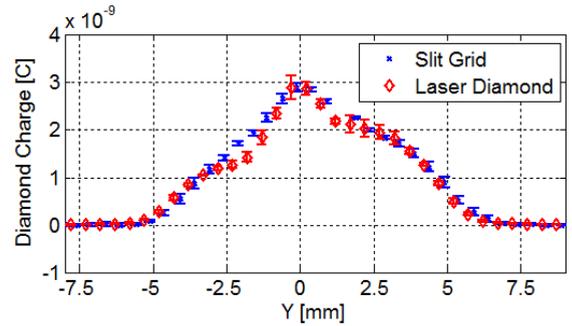

Figure 12. Comparison of beam profiles. Signal of the slit/grid signal normalized to the charge measured with the diamond detector.

Comparison of both phase space distributions shows a remarkable overall agreement. Minor differences are mainly present in the angular distribution. The measurement using the laserwire exhibits a slightly broadened angular profile, with the effect due to the lower special resolution of the diamond detector.

Integrating the phase space in the divergence domain, the resulting beam profile is plotted for both instruments in Figure . The disagreement in terms of profile shape and therefore extracted beam size is well below 2%. The error bars, calculated as the signal's amplitude variation in time along the LINAC4 pulse, are in the same range for both the laser/diamond and slit/grid systems.

### 1.7. *Emittance reconstruction*

The value of the normalized emittance can be expressed as follows:

$$\varepsilon_{RMS,norm} = \beta\gamma\sqrt{\langle y^2 \rangle \langle y'^2 \rangle - \langle yy' \rangle^2} \qquad (2)$$

Here y and y´ represent the amount of arriving particles at a given vertical position and divergence angle. The relativistic β and γ values allow transposition from the geometric to the normalized emittance.

The measurement depends strongly on the sampled phase space area due to the constant noise level of a real instrument. To suppress the noise a basic approach is used to exclude all values in phase space which are below a certain threshold corresponding to a percentage of the maximum amplitude. Figure  shows the resulting emittance values, measured with both the laser/diamond instrument as well as with the slit/grid reference system as a function of the applied threshold. The characteristic kink in this curve marks the spot where the noise is largely suppressed and the sampled signal starts to originate from impinging particles. For the laserwire system this point is quite distinct at 1.1%. The equivalent position for the slit/grid is not so clearly defined but is in the same region. Assuming the same threshold of 1.1% for both systems the resulting emittance values are summarized in Table 3 with the SD representing the uncertainty of the emittance measurements at intervals along the LINAC4 pulse.

Table 3. Comparison of emittance result values

| System | Norm. RMS Emittance | SD |
|---|---|---|
| Slit/Grid | 0.232 π mm mrad | 0.007 |
| Laser/Diamond | 0.239 π mm mrad | 0.0026 |

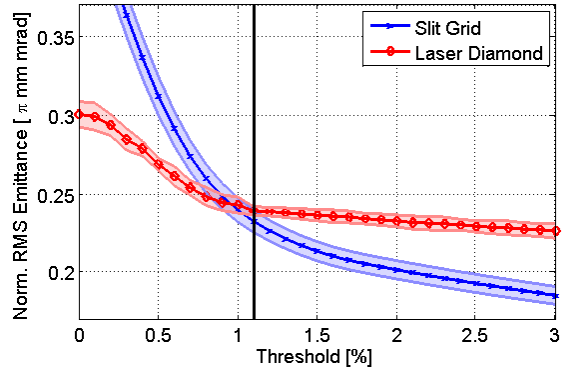

Figure 13. Normalized emittance value resulting from both instruments. Dependence on used threshold for noise suppression.

**Summary**

The principal design and first measurements of the novel laser-based instrument were already accomplished at the first commissioning stage of LINAC4 at 3 MeV [4], [5]. During the 12 MeV commissioning the design of instrument was refined and its performance was characterized in depth.

The amplitude of the diamond detector signal was validated by comparison with prior simulations. The linearity of the detector was characterized in the time domain along the LINAC4 pulse and also in comparison to the number of impinging particles. In both domains the detector showed a straight proportional behaviour. The homogeneity of the different detector channels also proved to be satisfactory. Overall the diamond detector system exhibited clearly improved performance due to the higher range of the impinging particles compared with the original 3 MeV tests.

With regard to the laser system, a noteworthy change in comparison to the 3 MeV setup is the extension of the fibre delivery of the laser beam to the IP to 10 m. The transported laser pulses did not show any distortion and the transmission efficiency was clearly above 70% even for pulses with peak power above 2 kW.

Analysing the obtained results recorded with the laserwire system, the temporal beam parameters as the time-of-flight or the bunch structure in the set of data was also found. The main focus of the measurements remained the profile and emittance reconstruction. The results were compared with the well-established slit/grid method. For both phase space and beam profile the agreement is outstanding. The RMS emittance value is strongly dependent on the threshold for noise suppression. However for a reasonable choice of threshold, the two systems agree to within 3%.

1.8. *Outlook*

The next steps of the LINAC4 commissioning at 50 MeV and 100 MeV will be used to test a modified version of the instrument, to measure the detached electrons and reconstruct the beam profile in a non-invasive manner [17]. Tests are ongoing to further increase the fibre delivery length of the laser beam in order to install the laser-head in a radiation safe environment outside the accelerator tunnel.

The encouraging results of the presented measurement campaign at 12 MeV represent a milestone on the way to the final instrument, which will be used at the 160 MeV LINAC4 beam. In preparation for its permanent operation, the electrode design of the diamond detector and its data acquisition readout chain will be re-designed to provide higher angular resolution and faster emittance measurements. The laser system will need to be modified to be able to scan in both the horizontal and vertical planes.

**Acknowledgements**

This paper is dedicated to the memory of our late colleague Christoph Gabor, a fruitful and enthusiastic scientist. We are grateful to the LINAC4 operation team and our colleagues in the CERN profile measurement section for the strong support during this project. Furthermore we want to thank the FETS project team for lending us their laser source and especially Alan Letchford and Jürgen Pozimski for the fruitful discussions. We acknowledge support from the STFS FETS grant and from the Marie Curie Networks LA³NET and oPAC which are funded by the European Commission under Grant Agreement Number 289191 and 289485.